\documentstyle[twocolumn,aps,prb]{revtex}
\begin{document}
\title{\hspace{7cm} {\em JETP Letters} {\bf 67} {\em (1998) 650--655}\\
\vskip 1cm
NONLINEAR SEEBECK EFFECT IN A MODEL GRANULAR SUPERCONDUCTOR}
\author{Sergei A. Sergeenkov}
\address{Bogoliubov Laboratory of
Theoretical Physics, Joint Institute for Nuclear Research\\
141980 Dubna, Moscow Region, Russia}
\address{\em (\today)}
\maketitle
\begin{abstract}
The change of the Josephson supercurrent density $j_s$ of a weakly-connected
granular superconductor in response to externally applied arbitrary
thermal gradient $\nabla T$ (nonlinear Seebeck effect) is considered within
a model of 3D Josephson junction arrays. For $\nabla T>(\nabla T)_c$,
where $(\nabla T)_c$ is estimated to be of the order of $\simeq 10^4K/m$ for
$YBCO$ ceramics with an average grain's size $d\simeq 10\mu m$, the
weak-links-dominated thermopower $S$ is predicted to become strongly
$\nabla T$-dependent.
\end{abstract}
\pacs{PACS numbers: 74.50.+r, 75.80.+q}

\narrowtext

A linear Seebeck effect, observed in conventional and high-$T_c$ ceramic
superconductors (HTS) and attributed to their weak-links structure (see,
e.g.,~\cite{ref1,ref2,ref3,ref4,ref5,ref6,ref7} and further references
therein), is based on the well-known fact that in a Josephson junction (JJ)
the superconducting phase difference $\Delta \phi$ depends only on the
supercurrent density $j_s$ (according to the Josephson relation
$j_s=j_c\sin \Delta \phi$,
where $j_c$ is the critical current density).
When a small enough temperature gradient $\nabla T$ is applied to such
a JJ (with the normal resistivity $\rho _n$),
the entropy-carrying normal current with density $j_n=S_0\nabla T/\rho _n$
is generated through such a junction, where
$S_0$ is the thermopower (a {\it linear} Seebeck coefficient). This normal
current density is locally cancelled by a counterflow of supercurrent with
density $j_s$, so that the total current density through the junction
$j=j_n+j_s=0$. As a result~\cite{ref3}, the supercurrent density
$j_s=-j_n$ generates a nonzero phase difference $\Delta \phi$
via a transient Seebeck thermoelectric field $E_T=\rho _nj_n=S_0\nabla T$
induced by the temperature gradient $\nabla T$.
If in addition, an external current of density $j_e$ also passes through the
weak link, a non-zero voltage will appear when the total current density
exceeds $j_c$, i.e., for $j_e=j_c\pm S_0\nabla T/\rho _n$.

In the present Letter, using a zero-temperature 3D model of Josephson
junction arrays, a nonlinear analog of the thermoelectric effect
(characterized by a non-trivial $\nabla T$-dependence of the
Seebeck coefficient $S$) in granular superconductors is considered.
The experimental conditions under which the predicted behavior of thermopower
can be observed in $YBCO$ ceramics are discussed.

The so-called $3D$ model of Josephson junction arrays (which is often used
to simulate a thermodynamic behavior of a real granular superconductor)
is based on the well-known tunneling Hamiltonian (see,
e.g.,~\cite{ref8,ref9,ref10,ref11,ref12,ref13})
\begin{equation}
{\cal H}(t)=\sum_{ij}^NJ_{ij}[1-\cos \phi _{ij}(t)],
\end{equation}
and describes a short-range interaction between $N$ superconducting grains
(with the gauge invariant phase difference $\phi _{ij}(t)$, see below),
arranged in a 3D lattice
with coordinates $\vec r_i=(x_i,y_i,z_i)$. The grains are separated by
insulating boundaries producing Josephson coupling $J_{ij}$ which is
assumed~\cite{ref8} to vary
exponentially with the distance $\vec r_{ij}$ between neighboring grains,
i.e., $J_{ij}(\vec r_{ij})=J(T)e^{-\vec \kappa \cdot \vec r_{ij}}$.
For periodic
and isotropic arrangement of identical grains (with spacing $d$ between
the centers of adjacent grains), we have
$\vec \kappa =(\frac{1}{d},\frac{1}{d},\frac{1}{d})$. Thus $d$ is of the
order of an average grain (or junction) size.

In general, the gauge invariant phase difference is defined as follows
\begin{equation}
\phi _{ij}(t)=\phi _{ij}(0)-A_{ij}(t),
\end{equation}
where $\phi _{ij}(0)=\phi _i-\phi _j$ with $\phi_i$ being the phase of the
superconducting order parameter, and $A_{ij}(t)$ is the so-called
frustration parameter, defined as
\begin{equation}
A_{ij}(t)=\frac{2\pi}{\Phi_ 0}\int_i^j\vec A(\vec r,t)\cdot d{\vec l},
\end{equation}
with $\vec A(\vec r,t)$ the (space-time dependent) electromagnetic vector
potential; $\Phi_ 0=h/2e$ is the quantum of flux, with $h$ Planck's constant,
and $e$ the electronic charge.

As is known~\cite{ref10,ref13}, a constant electric field $\vec E$ applied to
a single JJ causes a time evolution of the phase
difference. In this particular case Eq.(2) reads
$\phi _{ij}(t)=\phi _{ij}(0)+\omega _{ij}(\vec E)t$
where $\omega _{ij}(\vec E)=2e\vec E\cdot \vec r_{ij}/\hbar$ with
$\vec r_{ij}=\vec r_i-\vec r_j$ being the distance between grains.
If, in addition to the external electric field $\vec E$, the network of
superconducting grains is under the influence of an applied magnetic field
$\vec H$, the frustration parameter $A_{ij}(t)$ in Eq.(3) takes the
following form
\begin{equation}
A_{ij}(t)=\frac{\pi}{\Phi _0}(\vec H\wedge \vec R_{ij})\cdot \vec r_{ij}-
\frac{2\pi}{\Phi _0}\vec E\cdot \vec r_{ij}t.
\end{equation}
Here, $\vec R_{ij}=(\vec r_i+\vec r_j)/2$, and we have used the conventional
relationship between the vector potential $\vec A$ and (i) constant
magnetic field $\vec H=rot \vec A$ (with $\partial \vec H/\partial t=0$) and
(ii) homogeneous electric field $\vec E=-\partial \vec A/\partial t$ (with
$rot \vec E=0$).

There are at least two ways to incorporate a thermal gradient $\nabla T$
dependence into the above model. Namely, we can either invoke an analogy
with the above-discussed influence of an applied electric field on
the system of weakly-coupled superconducting grains or assume a direct
$\nabla T$ dependence
of the phase difference (as it was recently suggested by Guttman et
al~\cite{ref14}). For simplicity, in what follows we choose the first
possibility and assume an analogy with the conventional Seebeck effect.
Recall that application of a temperature
gradient $\nabla T$ to a granular sample is known to produce a
thermoelectric field~\cite{ref1,ref2} $\vec E_T=S_0\nabla T$, where $S_0$
is the so-called {\it linear} ($\nabla T$-{\it independent}) Seebeck
coefficient.
Assuming that in Eq.(4) $\vec E\equiv \vec E_T$, we arrive at the following
change of the junction phase difference under the influence of an applied
thermal gradient $\nabla T$
\begin{equation}
\phi _{ij}(t)=\phi _{ij}(0)-A_{ij}(t)
\end{equation}
with the frustration parameter
\begin{equation}
A_{ij}(t)=\frac{\pi}{\Phi _0}(\vec H\wedge \vec R_{ij})\cdot \vec r_{ij}-
\frac{2eS_0}{\hbar }\nabla T\cdot \vec r_{ij}t.
\end{equation}
As we see, the above equation explicitly introduces a direct $\nabla T$
dependence into the phase difference, expressing thus the main feature of
the so-called thermophase effect suggested by Guttman et al~\cite{ref14}.
Physically, it means that
the macroscopic normal thermoelectric voltage $V$ couples to the phase
difference on the junction through the quantum-mechanical Josephson
relation $V\propto d\Delta \phi /dt$.
Later on we shall obtain a rather simple connection between the
thermophase coefficient $S_T\equiv d\Delta \phi /d\Delta T$
and the conventional linear Seebeck coefficient $S_0$.

To consider a nonlinear analog of the Seebeck effect (characterized by a
$\nabla T$-dependent thermopower $S$), we recall~\cite{ref10,ref13} that
within
the model under consideration the supercurrent density operator $\vec j_s$
is related to the pair polarization operator $\vec p$ as follows ($V$ is a
sample's volume)
\begin{equation}
\vec j_s=\frac{1}{V}\frac{d\vec p}{dt}=
\frac{1}{i\hbar V}\left[ \vec p,{\cal H}\right ],
\end{equation}
where the polarization operator itself reads
\begin{equation}
\vec p=\sum_{i}^Nq_i \vec r_i.
\end{equation}
Here $q_i =-2en_i$ with $n_i$ the pair number operator, $r_i$ is the
coordinate of the center of the grain.

Finally, in view of Eqs.(1)-(8), and taking into account a usual
"phase-number" commutation relation, $[\phi _i,n_j]=i\delta _{ij}$, we find
\begin{equation}
<\vec j_s(\nabla T)>=\frac{2e}
{\hbar d^3} \sum_{ij}^N\int\limits_{0}^ {\tau }\frac{dt}{\tau}
\int\limits_{0}^{\infty}\frac{d\vec r_{ij}}{V}
J_{ij}(\vec r_{ij})\sin \phi _{ij}(t)\vec r_{ij}
\end{equation}
for the thermal gradient induced value of the averaged supercurrent density.
Here a temporal averaging (with a characteristic time $\tau$) accounts for
a change of the phase coherence during tunneling of Cooper pairs through the
barrier, while integration over the relative grain positions $\vec r_{ij}$
is performed bearing in mind a short-range character of the Josephson
coupling energy, viz. $J_{ij}(\vec r_{ij})=J(T)f(x_{ij})f(y_{ij})
f(z_{ij})$ with $f(u)=e^{-u/d}$.

To discuss a true $\nabla T$ induced thermophase effect only, in
what follows we completely ignore the effects due to a nonzero applied
magnetic field (by putting $\vec H=0$ in Eq.(6)) as well as rather important
in granular superconductors "self-field" effects (see ~\cite{ref12,ref13}
for discussion of this problem) and
assume that in equilibrium (initial) state
(with $\nabla T=0$) $<\vec j_s>\equiv 0$, implying thus $\phi _{ij}(0)
\equiv 0$. The latter condition in fact coincides with a current density
conservation requirement at zero temperature~\cite{ref9}.
As a result, we find that
an {\it arbitrary} temperature gradient $\nabla _xT\equiv \Delta T/\Delta x$,
applied along the $x$-axis to the Josephson junction network, induces the
appearance of the corresponding (nonlinear) longitudinal supercurrent
with density
\begin{equation}
j_s(\Delta T)\equiv <j_s^x(\nabla _xT)>=j_0G(\Delta T/\Delta T_0),
\end{equation}
where
\begin{equation}
G(z)=\frac{z}{1+z^2}.
\end{equation}
Here, $j_0=2eJNd/\hbar V$, $(\nabla _xT)_0\equiv \Delta T_0/\Delta x=
\hbar /2ed\tau S_0$, and $z=\Delta T/\Delta T_0$.

Notice that for a small enough temperature gradient (when $\nabla _xT\ll
\nabla _xT_0$), we recover a conventional {\it linear} Seebeck dependence
$j_s(\nabla _xT)=\alpha (T)S_0\nabla _xT$ with
$\alpha (T)=(2ed/\hbar )^2J(T)N\tau /V$.
On the other hand, for this result to be consistent with the above-discussed
conventional expression $j_s=S_0\nabla _xT/\rho _n$, zero temperature
coefficient $\alpha (0)$ should be simply related to the specific resistance
$\rho _n$. Let us show that this is indeed the case.
Using $J(0)=\hbar \Delta _0/4e^2R_n$ for zero-temperature Josephson energy
(where $\Delta _0$ is a zero-temperature gap parameter), $V\simeq Nld^2$
for sample's volume (with $l$ being a relevant sample's size),
and taking into account that the normal state resistance between grains
$R_n$ is related to $\rho _n$ as follows, $\rho _n\simeq (d^2/l)R_n$,
the self-consistency condition $\alpha (0)=1/\rho _n$ yields
$\tau \simeq (l/d)^2(\hbar /\Delta _0)$
for the characteristic Josephson time.

As it follows from Eq.(10),
above some threshold value $(\nabla _xT)_c\simeq 0.25(\nabla _xT)_0$ the
supercurrent density starts to substantially deviate from a linear law
suggesting thus the appearance of nonlinear Seebeck effect with
$\nabla T$-dependent coefficient $S(\nabla _xT)=S_0/(1+z^2)$ where
$z=\nabla _xT/(\nabla _xT)_0$ and $S_0\equiv S(0)$. Let us estimate an
order of magnitude of this threshold value of the thermal gradient needed to
observe the predicted nonlinear behavior of the thermopower in
weak-links-bearing HTS. Using $S_0 \simeq 0.5\mu V/K$ and
$\Delta _0/k_B \simeq 90K$ for
thermopower and zero-temperature gap parameter in $YBCO$,
and $l \simeq 0.5mm$ for a typical sample's size~\cite{ref2,ref4},
we get $\tau \simeq 10^{-9}s$ for the characteristic Josephson tunneling
time (Cf.~\cite{ref13}), and $(\nabla _xT)_c\simeq 10^4K/m$ for the threshold
thermal gradient in a granular sample with an average grain
(or junction) size $d\simeq 10\mu m$. Besides, taking $J(0)\simeq \Delta _0$
for a zero-temperature Josephson energy (in samples with
$R_n\simeq \hbar /4e^2$),
we arrive at the following reasonable estimate of
the weak-links-dominated critical current density
$j_0=2eJ/\hbar ld\simeq 10^3A/m^2$ in $YBCO$ ceramics. We believe that the
above estimates suggest quite an optimistic possibility to observe the
discussed nonlinear behavior of the thermoelectric power in (real or
artificially prepared) granular HTS materials and hope that the effects
predicted in the present paper will be challenged by experimentalists.

In conclusion, let us obtain the connection between the conventional
(linear) Seebeck effect and the above-mentioned thermophase effect
(which is linear by definition).
According to Guttman et al~\cite{ref14}, the latter effect is characterized
by a nonzero transport coefficient $S_T=d\Delta \phi/d\Delta T$.
In our particular case (with $\phi _{ij}(0)=0$ and $\vec H=0$), it follows
from Eqs.(5) and (6) that
\begin{equation}
\Delta \phi\equiv \frac{1}{\tau}\int_0^\tau dt\sum_{ij}
\frac{\phi _{ij}(t)}{N}\simeq \frac{e\tau S_0}{\hbar}\Delta T
\end{equation}
Hence, within our approach the above two $\nabla T$ induced {\it linear}
effects
(characterized by the transport coefficients $S_T$ and $S_0$, respectively)
are related to each other as follows
\begin{equation}
S_T\simeq \left (\frac{e\tau}{\hbar}\right )S_0\simeq
\left( \frac{e}{\Delta _0}\right )\left(\frac{l}{d}\right)^2S_0.
\end{equation}

To summarize, the change of the Josephson supercurrent density of a granular
superconductor under the influence of an arbitrary thermal gradient
(a nonlinear Seebeck effect) was considered within a model
of 3D Josephson junction arrays. A possibility of experimental observation
of the predicted effect in HTS ceramics was discussed.


\begin{thebibliography}{99}
\bibitem{ref1} R.P. Huebener, A.V. Ustinov, and V.K. Kaplunenko,
Phys. Rev. B {\bf 42}, 4831 (1990).
\bibitem{ref2} A.V. Ustinov, M. Hartman, and R.P. Huebener,
Europhys. Lett. {\bf 13}, 175 (1990).
\bibitem{ref3} A.V. Ustinov, M. Hartman, R.P. Huebener et al,
Supercond. Sci. Techn. {\bf 4}, S400 (1991).
\bibitem{ref4} R. Doyle and V. Gridin, Phys. Rev. B {\bf 45}, 10797 (1992).
\bibitem{ref5} R. Doyle and V. Gridin, Europhys. Lett. {\bf 19}, 423 (1992).
\bibitem{ref6} D.J. van Harlingen, Physica B {\bf 109-110}, 1710 (1982).
\bibitem{ref7}  G.I. Panaitov, V.V. Ryazanov, A.V. Ustinov et al,
Phys. Lett. A {\bf 100}, 301 (1984).
\bibitem{ref8} B. M\"uhlschlegel and D.L. Mills, Phys. Rev. B {\bf 29},
159 (1984).
\bibitem{ref9} C. Ebner and D. Stroud, Phys. Rev. B {\bf 31}, 165 (1985).
\bibitem{ref10} C. Lebeau, J. Rosenblatt, A. Robotou et al,
Europhys. Lett. {\bf 1}, 313 (1986).
\bibitem{ref11} V.M. Vinokur, L.B. Ioffe, A.I. Larkin et al,
ZhETF {\bf 93}, 343 (1987).
\bibitem{ref12} L. Leylekian, M. Ocio, L.A. Gurevich et al,
ZhETF {\bf 112}, 2079 (1997).
\bibitem{ref13}  S. Sergeenkov, J. Phys. I France {\bf 7}, 1175 (1997).
\bibitem{ref14}  G.D. Guttman, B. Nathanson, E. Ben-Jacob et al,
Phys. Rev. B {\bf 55}, 12691 (1997).
\end{thebibliography}
\end{document}